\documentclass[sigconf]{acmart}

\usepackage[inline]{enumitem}
\usepackage{tikz}
\usetikzlibrary{automata,arrows.meta,fit,positioning,chains,shapes.geometric,shapes.symbols,fit, matrix}

\AtBeginDocument{%
  }

\begin{document}

\setcopyright{cc}
\setcctype{by}
\acmConference{Doctoral Summer School in Immersive Media, Porto}
\acmYear{2026}
\acmMonth{5}

\title{Language Learning through Immersive Multi-User Gamification} 
\subtitle{DSSIM '26 --- Group 4: Lost in Translation}

\author{Rosalina Ribeiro Silva}
\email{rosalina.r.silva@inesctec.pt}
\orcid{0009-0005-3156-3839}
\affiliation{%
  \institution{University of Porto}
  \country{Portugal}
}

\author{Zichen Feng}
\email{zichen.feng@universite-paris-saclay.fr}
\affiliation{%
  \institution{Paris-Saclay University}
  \country{France}
}

\author{Rikard Olajos}
\email{rikard.olajos@cs.lth.se}
\orcid{0000-0002-5219-0729}
\affiliation{%
  \institution{Lund University}
  \country{Sweden}
}

\author{Tim Rolff}
\email{tim.rolff@uni-hamburg.de}
\orcid{0000-0001-9038-3196}
\affiliation{%
  \institution{University of Hamburg}
  \country{Germany}
}

\renewcommand{\shortauthors}{Group 4: Lost in Translation}

\iffalse
\newcommand{\Rikard}[1]{\textcolor{purple}{Rikard: #1}}
\newcommand{\Tim}[1]{\textcolor{orange}{Tim: #1}}
\newcommand{\Rosalina}[1]{\textcolor{blue}{Rosalina: #1}}
\else
\newcommand{\Rikard}[1]{}
\newcommand{\Tim}[1]{}
\newcommand{\Rosalina}[1]{}
\fi

\begin{abstract}
We propose an immersive virtual-reality game designed to enable multi-user, multi-language learning during co-play. We hypothesize that our method will be more engaging, leading to better rhythm, pronunciation, memorization, and speaking confidence. We implement the experience as a fantasy role-playing game by introducing vocabulary and phrases to the user as spells to learn. Players then cast more efficient spells depending on their language skills.
For evaluation, we compare the game to a single-user baseline that replicates traditional language learning in a non-immersive setting. We measure short- and long-term retention of vocabulary and phrases, as well as scoring rhythm and pronunciation.
Furthermore, we collect user data on engagement and mental workload.
\end{abstract}




\begin{teaserfigure}
  \begin{center}
  \includegraphics[width=0.92\textwidth]{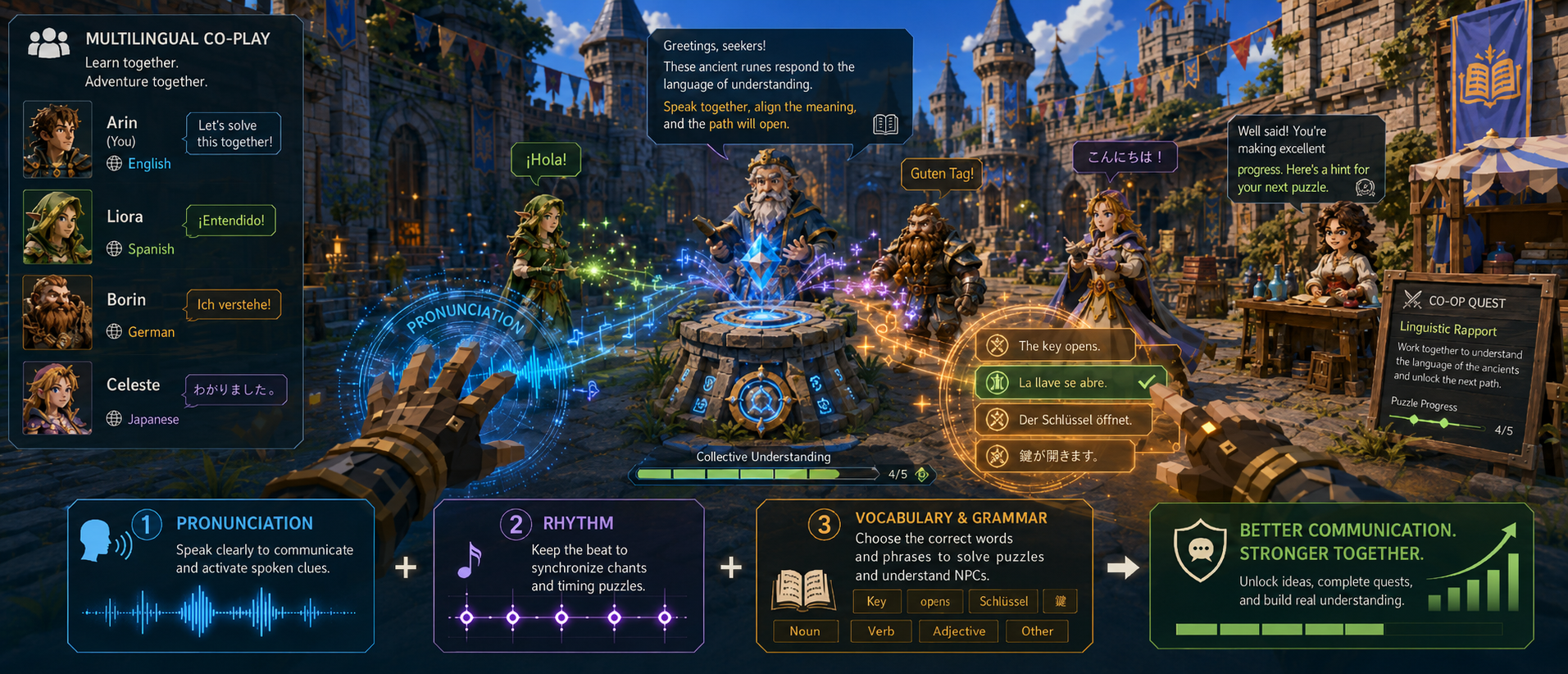}
  \end{center}
  \caption{Concept art of our proposed language learning application. The application should support analysis of pronunciation, rhythm, vocabulary, and grammar. We aim at gamifying these elements by scaling gameplay elements with language proficiency. (Image generated with ChatGPT.)}
  \Description{TODO: Long text description of the teaser figure for accessibility.}
  \label{fig:teaser}
\end{teaserfigure}

\maketitle
%
%

\section{Introduction and Problem Definition}
Learning a language often requires more than just memorizing vocabulary \cite{canale1980theoretical,moir2002learners}. Learners must develop skills for pronunciation \cite{munro2006mutual,loewen2017pronunciation}, rhythm \cite{celce2010teaching}, intonation \cite{ramirez2021intonation}, listening skills \cite{vandergrift20041}, and confidence in speaking \cite{macintyre1998conceptualizing}. Besides, an often overlooked part of learning a language are the manners and cultural customs of native speakers \cite{kramsch1993context}. However, a lot of different learning applications and platforms focus on either one of these or a small subset \cite{ishaq2021mobile,lim2024apps,rosell2017state}.
While some applications, such as Duolingo\footnote{\url{https://en.duolingo.com}}, PolyChat\footnote{\url{https://www.polychatapp.com}}, or LangoTalk\footnote{\url{https://www.langotalk.org}} already utilize large language models (LLMs) to teach common language skills. This is often done in a gamified application, which has been shown to improve language learning \cite{deterding2011game,luo2023effectiveness,sadeghi2022effects,ishaq2021mobile}. Yet, these applications are only for one-on-one interactions. This is in contrast with language learning research, which indicates that people are often learning better in a group~\cite{philp2013peer,long1996role,alzubi2024examining}.

This raises the fundamental challenge that people of one predefined group might not want to learn the same language but rather different ones but still want to spend time together or are more at ease trying to learn a new language with people they know. In a classroom setting, this is almost impossible to arrange, as it would require different language teachers, again leading to individual teaching defeating the purpose and benefits of learning in a group. It might also be the case that the people in these groups are geographically separated and can therefore not physically learn together. Even though some language learning apps, such as Duolingo, include social network features and support a set of skills necessary for language learners \cite{ishaq2021mobile}, they still provide very rudimentary ways to interact with each other when learning different languages.

Beyond these language-learning fundamentals, immersive virtual reality (VR) offers additional opportunities for situated and contextualized language practice. Previous reviews indicate that VR has been explored for a range of language-learning activities, including interactive communication, task-based instruction, vocabulary learning, speaking, listening, cultural knowledge, and learner motivation \cite{lin2015language,parmaxi2023virtual,hua2023virtual}.\\


Despite the previously mentioned applications tackling the already mentioned challenges, the challenge of multi-language learning in a group setting still persists as a research gap.
Therefore, in this paper, we argue that a gamified VR application, which includes the concepts of memorization, pronunciation, and rhythm inside a gamified immersive VR application while simultaneously providing multi-user language learning, improves the user's language skills. To investigate how well such a system improves language learning for users, we came up with the following research questions (RQs):
\begin{description}
\item[\textbf{RQ1}.] How does multilingual co-play, in which players learn different target languages within the same game session, affect language-learning outcomes and collaborative engagement?

\item[\textbf{RQ2}.] How does integrating pronunciation, rhythm, grammar, and vocabulary into core gameplay mechanics affect players' language-learning performance compared to non-gamified practice?


\end{description}


\section{Related work}
In this section, we will briefly cover the topics of second language learning through the lens of gamification and VR. We will also concisely touch on the subject of LLMs as tools for language learning and the concept of multiple people learning a different language while being co-located in the same application. 



\subsection{Gamification in Second Language Learning}
Research indicates that gamification offers another important aspect to make language practice more engaging and persistent \cite{deterding2011game,luo2023effectiveness,sadeghi2022effects,ishaq2021mobile}. 
In educational contexts, game elements such as goals, progress indicators, rewards, feedback, and challenges can support motivation and engagement when they are meaningfully aligned with the learning activity rather than added as superficial incentives \cite{deterding2011game,toda2019analysing}. This is particularly relevant for language learning, where learners benefit from repeated practice, immediate feedback, and opportunities to use language in context \cite{sykes2012language,reinhardt2026digital}. Empirical work on gamified foreign-language learning further suggests that game-based or gamified activities can support vocabulary learning and learner motivation \cite{sadeghi2022effects,luo2023effectiveness}.

Games also provide a structure in which language-learning tasks can be embedded into meaningful actions. Rather than treating vocabulary, grammar, pronunciation, and rhythm as isolated exercises, game mechanics can connect them to goals, quests, interaction, and progression. This is consistent with perspectives on digital games as language-learning environments that support interaction, feedback, motivation, and contextualized practice \cite{sykes2012language,reinhardt2026digital}. For example, vocabulary can be discovered through exploration, grammar can be practiced through quest dialogue, pronunciation can affect spoken interaction, and rhythm or prosody can become part of timed challenges. By linking language skills to in-game abilities and progression, the learning activity becomes part of the core gameplay rather than an external exercise layer.

\begin{figure*}[!t]
    \centering
    \resizebox{0.9\textwidth}{!}{
        \begin{tikzpicture}
  \tikzset{
    bound/.style={
      draw,
      minimum height=2cm,
      inner sep=1em,
    },
    arrow/.style={
      minimum height=1.7cm,
      inner sep=1em,
      xshift=0.5em,
      shape=signal,
      signal from=west,
      signal to=east,
      signal pointer angle=110,
    }
  }
  \begin{scope}[start chain=transition going right,node distance=-\pgflinewidth]
    \node[arrow,on chain,align=center, fill=orange!20] {Consent form\\and signature};
    \node[arrow,on chain,align=center, fill=orange!25] {Fill out\\demographics\\ questionnaire};
    \node[arrow,on chain,align=center, fill=orange!30] {Select\\ language\\from list};
    \node[arrow,on chain,align=center, fill=orange!35] {Fill out study\\questionnairs\\(NASA-TLX, ...)};
    \node[arrow,on chain,align=center, fill=orange!40] {Test\\language\\proficiency};
  \end{scope}

  \begin{scope}[yshift=-2cm,start chain=transition going right,node distance=-\pgflinewidth]
    \node[arrow,on chain,align=center, fill=orange!45] {Start\\tutorial};
    \node[arrow,on chain,align=center, fill=orange!50] {Learn\\language};
    \node[arrow,on chain,align=center, fill=orange!55] {Perform\\quests};
    \node[arrow,on chain,align=center, fill=orange!60] {Language\\ proficiency\\test};
    \node[arrow,on chain,align=center, fill=orange!65] {Fill out study\\questionnairs\\(NASA-TLX, ...)};
    \node[arrow,inner sep=0em,on chain,align=center, fill=orange!70] { };
    \node[arrow,on chain,align=center, fill=orange!75] {Long-term\\language\\proficency test};
  \end{scope}
\end{tikzpicture}
    }
    \caption{Predefined study procedure when running our VR study.}
    \label{fig:procedure}
\end{figure*}
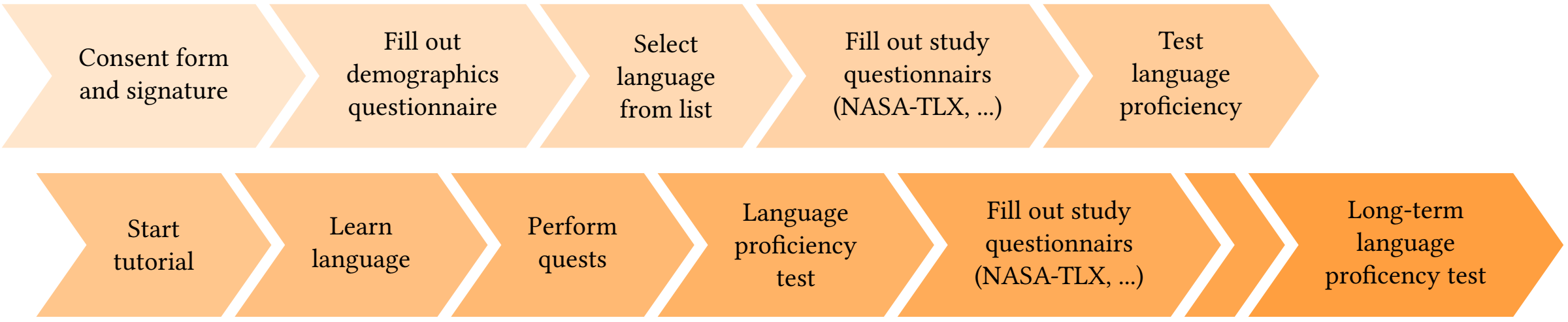
    
\subsection{VR for Second Language Learning}
Unlike conventional screen-based applications, VR can place learners inside interactive scenarios in which language use is connected to objects, characters, places, and actions in the environment \cite{slater2009place,dalgarno2010learning,santos2016augmented}. This can make language practice more contextualized and meaningful, as learners do not only repeat words but use them to act within a simulated situation \cite{dede2009immersive,dalgarno2010learning}.
Immersive VR may therefore be particularly useful for practicing communicative language use. Systematic reviews suggest positive effects of immersive VR on foreign-language learning, motivation, and satisfaction, while also emphasizing that the evidence base is still developing and that careful pedagogical design is necessary \cite{parmaxi2023virtual,peixoto2021immersive,hua2023virtual}.

Empirical work further supports the potential of immersive VR for specific language-learning tasks. For example, \citet{legault2019immersive} found that learners showed higher recognition accuracy for Mandarin words learned in immersive VR compared to words learned through word (i.e., word-paired association) \cite{legault2019immersive}. However, VR should not be understood as beneficial simply because it is immersive. Its value depends on whether the virtual environment~(VE) creates meaningful, interactive, and pedagogically grounded opportunities for language use \cite{parmaxi2023virtual}.

\subsection{LLMs for Second Language Learning}
The recent advancements in deep learning, especially in LLMs, have renewed the research interest in applying artificial intelligence (AI) for second language learning. Their ability of LLMs to produce open-ended language, maintain dialog, understand tasks, adapt to learners input, and provide feedback across a wide array of linguistic tasks \cite{kohnke2023chatgpt,yang2024chatgpt,lee2026generative} provides enormous potential for language-learning activities. Especially as LLMs are trained on massive corpora of different languages \cite{grattafiori2024llama,radford2018improving}, they implicitly learn the structure, grammar, and common phrases of that language \cite{radford2018improving}.
This allows scenarios that require interaction, variation, and personalization, including grammar support, vocabulary explanation, writing feedback, role-play, and scenario-based dialog. 

However, the usage of LLMs still raises pedagogical and technical challenges. Current models may produce factually wrong, inappropriate, overly generic, or misleading output \cite{barrot2023using,yoon2023evaluation,han2024chatgpt}, especially when used without teacher oversight or task-specific design. For language learning in games, LLMs are of particular relevance, as they support flexible systems for quests and conversation \cite{karaosmanoglu2024language}. 


\section{Conceptual Approach}
\label{sec:conceptual approach}

To answer our RQs, we propose a multi-user, multi-language VR experience that users can interactively explore and solve challenges in a fantasy role-playing game (RPG) while being fully immersed in the VR experience. In the context of our system, VR is not only used as a visualization technology but also as a way to situate pronunciation, rhythm, vocabulary, and grammar practice inside gamified collaborative fantasy-role-playing activities.

We chose a fantasy scenario to avoid biasing participants through a preselected cultural exhibition, reducing the confounds during the evaluation, but also because research indicates that fantasy elements improve language learning \cite{araminta2020role}. The challenges, game mechanics, and quests will be designed to force users to use and exercise vocabulary, phrases, and pronunciation in a language of their choosing. For example, we gamify learning through the fantasy RPG experience, requiring users to cast spells with the language of their choosing. We also expect that the gamification aspects increase the engagement \cite{sailer2020gamification,hamari2014does} and, at the same time, give us a platform for naturally incorporating multi-user interactions \cite{peterson2010massively,peterson2012learner}. To perform speech pattern analysis, we rely on systems that commonly analyze learner speech at different levels, including phoneme-, word-, and utterance-level accuracy, and increasingly also consider features such as stress, timing, pauses, intonation, and rhythm \cite{ngo2024effectiveness}. Here, automatic pronunciation assessment and computer-assisted pronunciation training can provide learners with timely feedback on spoken language production \cite{witt2000phone,neumeyer2000automatic,el2023automatic}.

By creating a sense of presence (i.e., the illusion of being there \cite{slater2018immersion}), VR can simulate situations in which learners need to speak, listen, respond, and make decisions in the target language. For example, learners may order food, ask for directions, solve a quest, or interact with virtual agents in a foreign-language. Such scenarios can provide opportunities for repeated practice while reducing some logistical barriers of real-world immersion, such as travel, access to native speakers, or fear of making mistakes in public.

\subsection{Hypotheses}
To answer our research questions given our conceptual approach and traditional language learning as a baseline, we divide our hypotheses into three different categories and state them as follows:\\

\noindent \textbf{Motivation}:
\begin{description}
    \item[H1.] People enjoy language learning in a group setting more than individually.
    \item[H2.] Participants stay more engaged with the immersive VR environment than with the baseline.
\end{description}

\noindent \textbf{Proficiency}:
\begin{description}
    \item[H3.] Immediately after the game, people will have an improved language proficiency and feel more confident speaking.
    \item[H4.] Gamification improves the language proficiency, rhythm, and pronunciation of participants compared to a non-gamified language learning.
\end{description}

\noindent \textbf{Long-term retention}:
\begin{description}
    \item[H5.] People who used VR remembered more words after a longer time than the baseline.
    \item[H6.] Cognitive load is higher in the VR condition than in the baseline, but engagement is higher in the VR condition, leading to better word retention over time.
    \item[H7.] People perform better on rhythm and pronunciation compared to baseline time.
\end{description}

\section{Reflection on Feasibility and Evaluation}

\subsection{Methodology}
\label{sec:methodology}
\begin{figure}
    \centering
    \begin{tikzpicture}[
block/.style={inner sep=6pt, text centered},
u1/.style={fill=purple!30},
u2/.style={fill=yellow!40},
]

\node[] (start1) at (0,0.6) {Quest Start};
\node[] (start2) at (3,0.6) {Quest Start};
\node[block, u1] (user1) at (0,0) {User 1};
\node[block, u2] (user2) at (3,0) {User 2};
\node[draw=green!30!black!60, thick, fit={(start1) (user1)}] (container11) {};
\node[draw=green!30!black!60, thick, fit={(start2) (user2)}] (container12) {};

\node[] (learn1) at (0,-1.4) {Spell Learning};
\node[] (learn2) at (3,-1.4) {Spell Learning};
\node[block, u1] (spell1) at (0,-2) {in French};
\node[block, u2] (spell2) at (3,-2) {in German};
\node[draw=red!70!black!60, thick, fit={(learn1) (spell1) }] (container21) {};
\node[draw=red!70!black!60, thick, fit={(learn2) (spell2)}] (container22) {};

\node[] (solve1) at (0,-3.4) {Quest Solving};
\node[] (solve2) at (3,-3.4) {Quest Solving};
\node[minimum height=21pt, trapezium, trapezium left angle=0, trapezium right angle=0, u1] (challenge11) at (0,-4) {Challenge 1};
\node[minimum height=21pt, trapezium, trapezium left angle=0, trapezium right angle=0, u2] (challenge12) at (3,-4) {Challenge 1};

\node[minimum height=21pt, trapezium, trapezium left angle=0, trapezium right angle=0, u1] (challenge21) at (0,-5) {Challenge 2};
\node[minimum height=21pt, trapezium, trapezium left angle=0, trapezium right angle=0, u2] (challenge22) at (3,-5) {Challenge 2};

\node[minimum height=21pt, trapezium, trapezium left angle=0, trapezium right angle=0, u1] (challenge31) at (0,-6) {Challenge 3};
\node[minimum height=21pt, trapezium, trapezium left angle=0, trapezium right angle=0, u2] (challenge32) at (3,-6) {Challenge 3};

\node[draw=blue!40!black!60, thick, fit={(solve1) (challenge11) (challenge21) (challenge31)}] (container31) {};
\node[draw=blue!40!black!60, thick, fit={(solve2) (challenge12) (challenge22) (challenge32)}] (container32) {};

\draw[-stealth] (container11) -- (container21);
\draw[-stealth] (container12) -- (container22);
\draw[-stealth] (container21) -- (container31);
\draw[-stealth] (container22) -- (container32);

\end{tikzpicture}
    \caption{Baseline VR experience where users do not interact to solve quests.}
    \Description{}
    \label{fig:single-user_quest}
\end{figure}
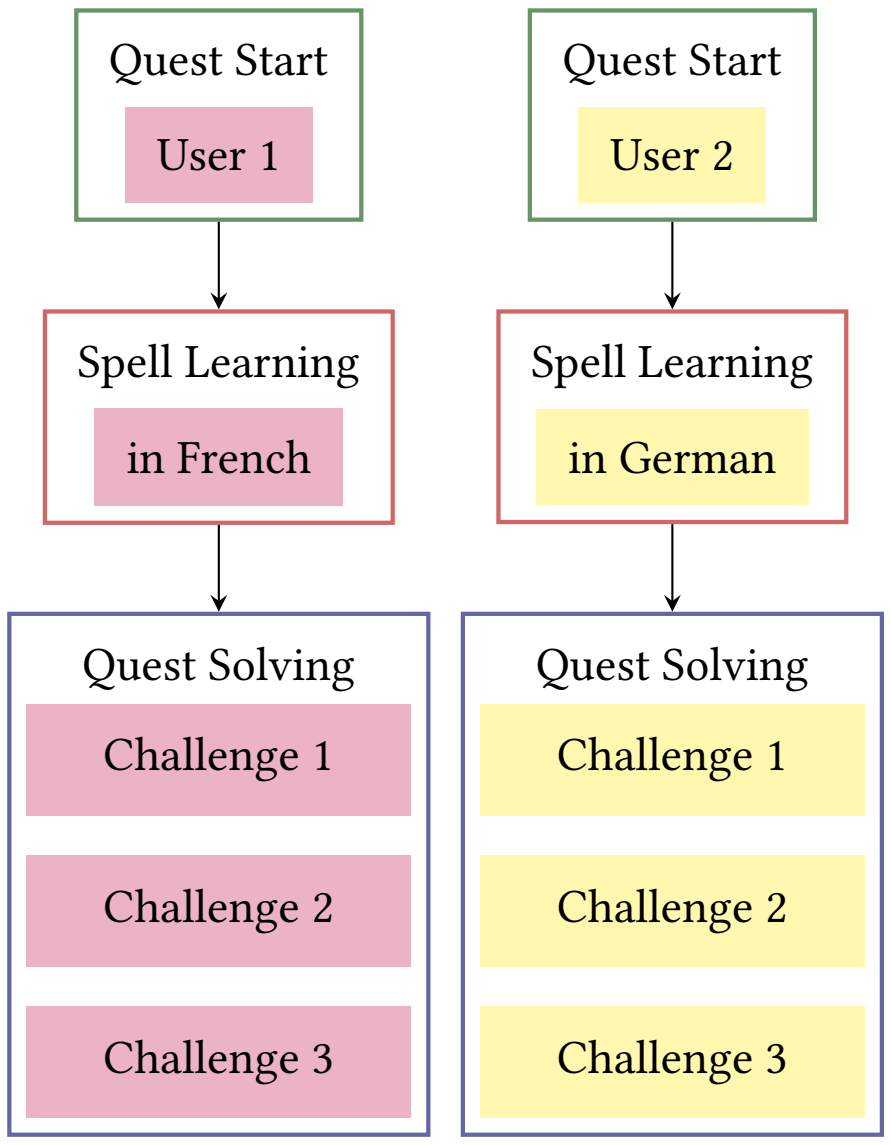

To test our hypotheses, we propose a between-subject case study with two conditions (baseline vs. application) that evaluates our approach to a reference group. This way we can analyze the language proficiency gained by our method. The experiment will be carried out in three phases:

\begin{enumerate*}[label=(\roman*)]
\item The first phase involves developing the VR application as a co-design study in consultation with language experts.
We invite language teachers to help us identify vocabulary and phrases that are suitable for a beginner to learn and discuss what is the most pedagogical way to teach it.
These experiences go into designing the spells in the fantasy RPG as well as the test for the baseline learners.\\

\item In the second phase, we establish the baseline condition.
Test participants are asked to fill in a pre-experiment assessment form to ascertain their current proficiency in a language.
The test participants then get to study phrases and vocabulary that were constructed in the first phase.
The experiment is implemented as a web-based study with a traditional flip card and listen-and-repeat language learning approach.
At the end of the experiment, the test participants are evaluated on how much of the language they retained.
Their pronunciation and rhythm are also judged and scored through pronunciation scoring \cite{witt2000phone,neumeyer2000automatic,el2023automatic}.
After four weeks, the same test participants are asked to redo the evaluation, and long-term retention is assessed.
Four-week cycles has been used in previous work for assessing retention of language learning~\cite{rogers2020retention,rogers2021retention}.
\\

\item In the final phase, we invite groups to establish the application condition. The participants that partake in a VR experience will not overlap with the baseline from the baseline condition (cf.~\autoref{sec:study design} for more details).
The groups are subjected to the same pre-experiment assessment as in the second phase before they carry out the VR experience.
To compare a multi-user immersive experience to a single-user immersive experience, a second configuration of the fantasy quest is presented to half of the participants.
This second configuration is not allowing the users to interact with each other, and the spell learning and quest solving is done in isolation (see Figure~\ref{fig:single-user_quest} and compare with Figure~\ref{fig:multi-user_quest}).
\end{enumerate*}

\begin{figure}
    \centering
    \resizebox{\columnwidth}{!}{%
    \begin{tikzpicture}[
phase/.style={signal, signal from=nowhere, inner sep=8pt, text centered, rotate=270},
desc/.style={rectangle, inner sep=5pt, anchor=west},
]

\node[phase, signal to=east, fill=orange!30] (phase1) {Development};
\node[phase, signal from=west, signal to=east, fill=green!50!black!30, anchor=west, xshift=2pt] (phase2) at (phase1.east) {Baseline study};
\node[phase, signal from=west, signal to=east, fill=blue!50!red!30, anchor=west, xshift=2pt] (phase3) at (phase2.east) {Multi-User VR study};

\node[desc, right of=phase1, xshift=16pt, yshift=-20pt] (desc1) at (phase1.west) {\textbf{Phase I}};
\node[desc, yshift=-15pt] (desc11) at (desc1.west) {$\bullet$ Co-design of the application and experiment with language experts};
\node[desc, yshift=-15pt] (desc12) at (desc11.west) {$\bullet$ Identify words and phrases for different languages to learn};
\node[] (right1) at (11, 0) {};
\node[draw=brown, thick, fit={(desc1) (desc11) (desc12) (right1)}, inner sep=4pt]  (box1) {};

\node[desc, right of=phase1, xshift=18pt, yshift=-6pt] (desc2) at (phase2.west) {\textbf{Phase II}};
\node[desc, yshift=-15pt] (desc21) at (desc2.west) {$\bullet$ Pre-experiment assessment of subjects};
\node[desc, yshift=-15pt] (desc22) at (desc21.west) {$\bullet$ Baseline study as a web-based experiment};
\node[desc, yshift=-15pt] (desc23) at (desc22.west) {$\bullet$ Let participants read out rhythms, pronunciations, words, and phrases};
\node[desc, yshift=-15pt] (desc24) at (desc23.west) {$\bullet$ Re-evalute participants after a week to measure long-term retention};
\node[] (right2) at (11, -3) {};
\node[draw=green!50!black!70, thick, fit={(desc2) (desc21) (desc22) (desc23) (desc24) (right2)}, inner sep=4pt]  (box2) {};

\node[desc, right of=phase1, xshift=20pt, yshift=-16pt] (desc3) at (phase3.west) {\textbf{Phase III}};
\node[desc, yshift=-15pt] (desc31) at (desc3.west) {$\bullet$ Invite groups of subjects to the VR experiment};
\node[desc, yshift=-15pt] (desc32) at (desc31.west) {$\bullet$ Pre-experiment assessment of subjects};
\node[desc, yshift=-15pt] (desc33) at (desc32.west) {$\bullet$ Participants partake in single- or multi-user VR experience};
\node[desc, yshift=-15pt] (desc34) at (desc33.west) {$\bullet$ Re-invite participants after four weeks to measure long-term retention};
\node[] (right3) at (11, -6) {};
\node[draw=blue!50!red!70, thick, fit={(desc3) (desc31) (desc32) (desc33) (desc34) (right3)}, inner sep=4pt]  (box3) {};

\end{tikzpicture}
    }
    \caption{Our proposed experimental setup.}
    \Description{}
    \label{fig:method}
\end{figure}
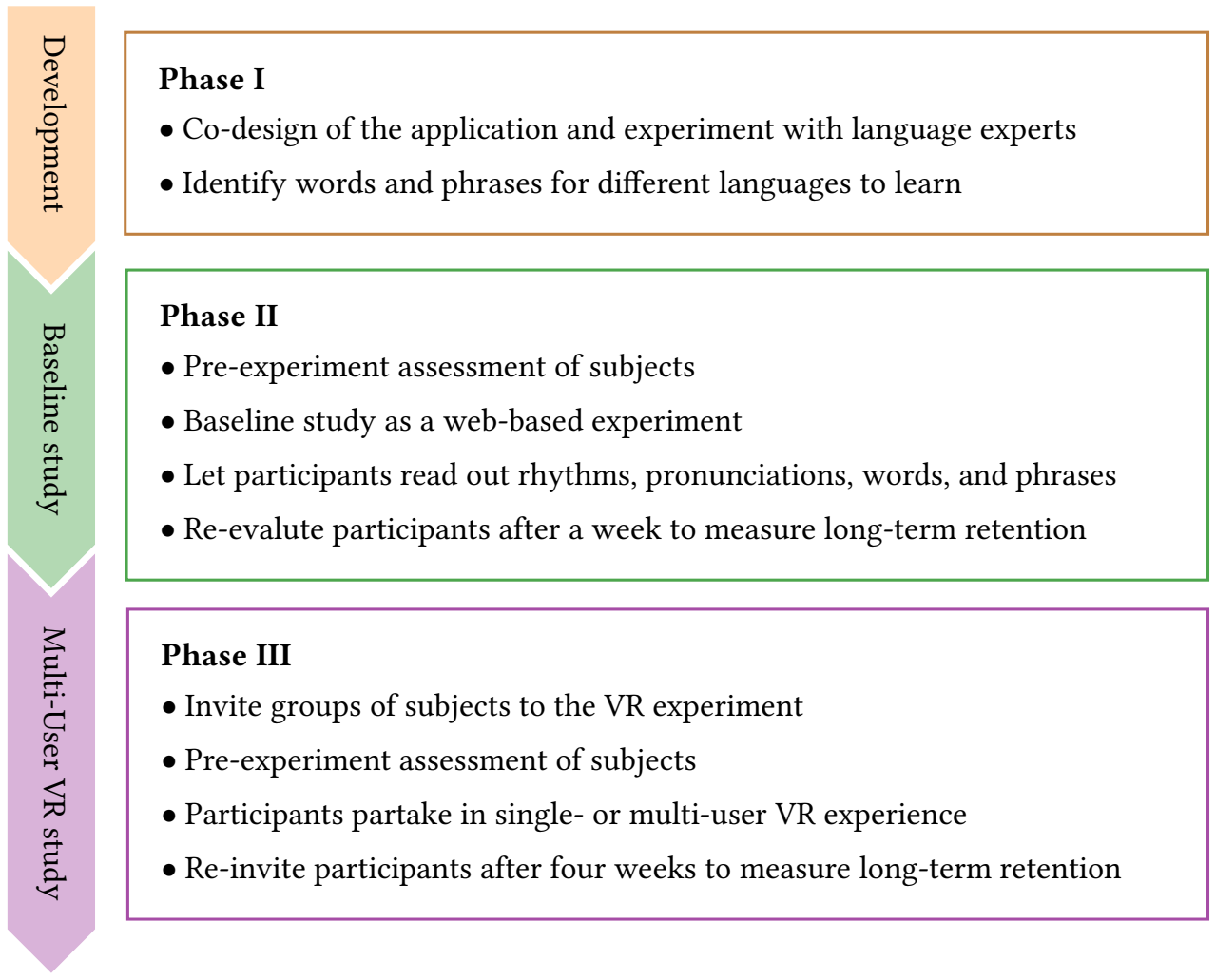

The single- and multi-user VR experiences evaluate the participants through the quest solving where the spells and challenges expose the users to the same vocabulary and phrases as the baseline group in the second phase.
The same pronunciation scoring software used on the baseline group is incorporated in the VR experience.
The full experiment is summarized in Figure~\ref{fig:method}.

\subsection{VR Study Procedure}
\label{sec:study design}
\Tim{This is currently a bit redundant with what \autoref{sec:methodology} does. Maybe its good to reduce content later.}
To evaluate our RQs and hypotheses, we designed our study procedure
as shown in \autoref{fig:procedure}. At the start of the session, we inform participants about the study and ask for their informed consent, requiring the signature of the participant. Next, we provide each participant with a unique identification and ask them to fill out a demographics questionnaire.
Then we let the participants indicate their preferred language to learn and we restrict participants to a language they will have little to no prior knowledge of.
This ensures that we do not have an additional variable from users with different language skills during our evaluation.
We will retain the right to pick a language for the participant if needed as we need a mix of languages in a learning session.
To evaluate the initial language proficiency of participants, we provide them with a language proficiency test of their chosen language.

After this setup period, each participant group is introduced to the application with the task to solve a fantasy RPG-themed quest. The quest task includes solving some challenges that will require all users to use words and phrases of their chosen language to progress. To ensure that all participants are familiar with the controls, we provide a short tutorial where participants are instructed to seek teachers who are embodied virtual agents represented as wizards, sorcerers, or witches. These agents teach the participants helpful spells that will be used in the forthcoming challenges. 
Here, the users learn and practice their spells while getting feedback on the accuracy of pronunciation, rhythm, and vocabulary. To link the language learning to game mechanics, we link the level of pronunciation, rhythm, fluency, and vocabulary retention back into the game by scaling the power of a spell with the participant's language proficiency in these categories. In case a participant forgets a language skill or wants to practice more, they can always go back to the language teacher.

\begin{figure}
    \centering
    \begin{tikzpicture}[
block/.style={inner sep=6pt, text centered},
u1/.style={fill=purple!30},
u2/.style={fill=yellow!40},
u3/.style={fill=blue!60!red!40},
]

\node[] (start) at (2,0.6) {Quest Start};
\node[block, u1] (user1) at (0,0) {User 1};
\node[block, u2] (user2) at (2,0) {User 2};
\node[block, u3] (user3) at (4,0) {User 3};
\node[draw=green!30!black!60, thick, fit={(start) (user1) (user2) (user3)}] (container1) {};

\node[] (learn) at (2,-1.4) {Spell Learning};
\node[block, u1] (spell1) at (0,-2) {in French};
\node[block, u2] (spell2) at (2,-2) {in German};
\node[block, u3] (spell3) at (4,-2) {in Spanish};
\node[draw=red!70!black!60, thick, fit={(learn) (spell1) (spell2) (spell3)}] (container2) {};

\node[] (solve) at (2,-3.4) {Quest Solving};
\node[minimum height=21pt, trapezium, trapezium left angle=0, trapezium right angle=-45, u1] (challenge11) at (0.55,-4) {};
\node[minimum height=21pt, trapezium, trapezium left angle=45, trapezium right angle=0, u3] (challenge13) at (3.45,-4) {};
\node[minimum height=21pt, trapezium, trapezium left angle=45, trapezium right angle=-45, u2] (challenge12) at (2,-4) {};
\node[fill=white, opacity=0.6, text opacity=1.0] (challenge1label) at (2,-4.02) {\Large Challenge 1};

\node[minimum height=21pt, trapezium, trapezium left angle=0, trapezium right angle=-45, u2] (challenge21) at (0.55,-5) {};
\node[minimum height=21pt, trapezium, trapezium left angle=45, trapezium right angle=0, u1] (challenge23) at (3.45,-5) {};
\node[minimum height=21pt, trapezium, trapezium left angle=45, trapezium right angle=-45, u3] (challenge22) at (2,-5) {};
\node[fill=white, opacity=0.6, text opacity=1.0] (challenge2label) at (2,-5.02) {\Large Challenge 2};

\node[minimum height=21pt, trapezium, trapezium left angle=0, trapezium right angle=-45, u3] (challenge31) at (0.55,-6) {};
\node[minimum height=21pt, trapezium, trapezium left angle=45, trapezium right angle=0, u2] (challenge33) at (3.45,-6) {};
\node[minimum height=21pt, trapezium, trapezium left angle=45, trapezium right angle=-45, u1] (challenge32) at (2,-6) {};
\node[fill=white, opacity=0.8, text opacity=1.0] (challenge3label) at (2,-6.02) {\Large Challenge 3};

\node[draw=blue!40!black!60, thick, fit={(solve) (challenge11) (challenge12) (challenge13) (challenge21) (challenge22) (challenge23) (challenge31) (challenge32) (challenge33)}] (container3) {};

\draw[-stealth] (container1) -- (container2);
\draw[-stealth] (container2) -- (container3);

\end{tikzpicture}
    \caption{The VR experience is centred around a multi-user fantasy role-playing quest.}
    \Description{}
    \label{fig:multi-user_quest}
\end{figure}
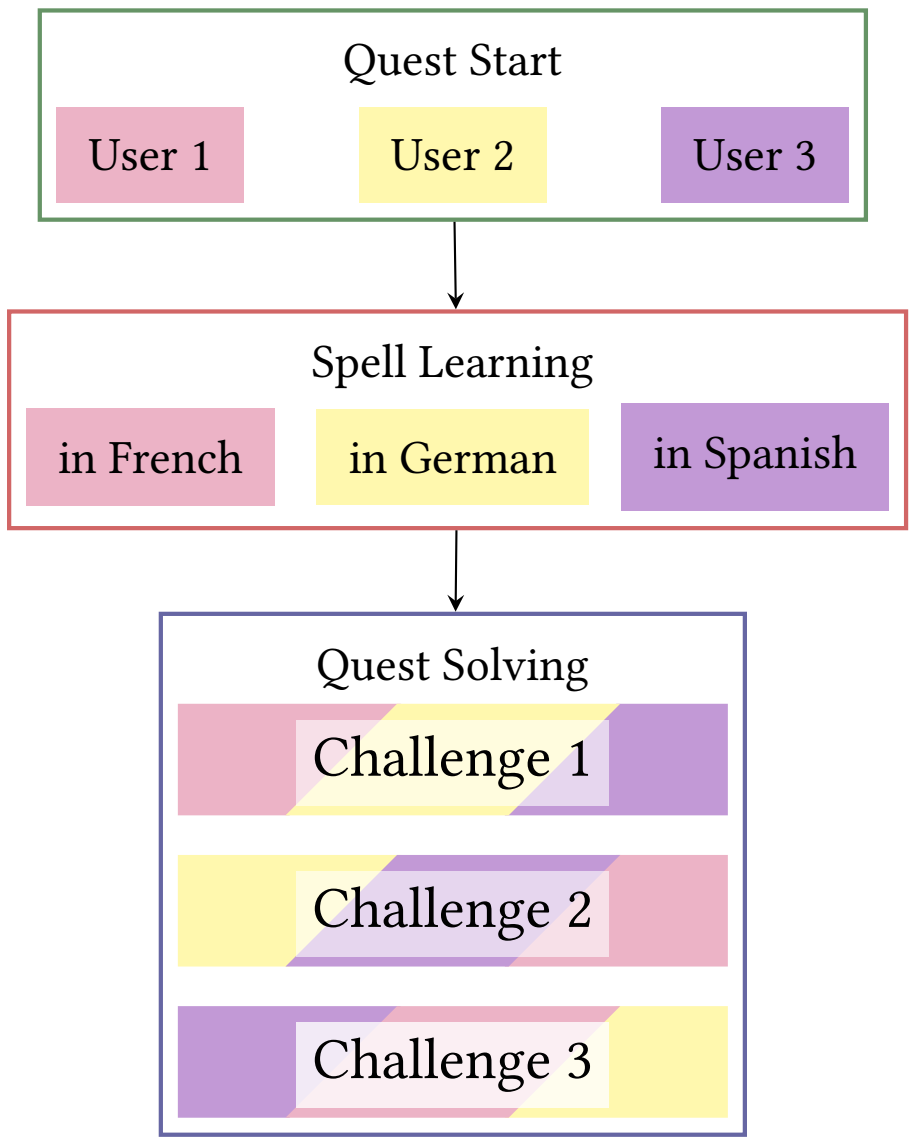

We then let players solve a series of quests that are given by different virtual agents. Here, participants are required to cooperate to solve challenges by trying to understand the virtual agent and by recalling their respective spells. We assume that people share the capability of one common language and will provide quest instructions in that language except for the learned words. During a quest, each challenge can combine the spells in different manners, which will make the users interact and help each other to progress, solving puzzles and riddles. An example of how the layout of a quest could look is presented in Figure~\ref{fig:multi-user_quest}.

\subsection{Proposed Analysis}
To answer our hypotheses, we will use a user feedback questionnaire, asking participants if they enjoyed learning in a group setting more than individually.
\textbf{H1} will be answered by the User Experience Questionnaire (UEQ)~\cite{schrepp2008ueq}.
\textbf{H2} will be answered by the Game Engagement Questionnaire (GEQ) \cite{brockmyer2009development}. For \textbf{H3}, we let participants perform another language proficiency test. Analogously, we will perform a long-term retention test after four weeks to answer \textbf{H5}. We will correlate the game score and the measured language proficiency to answer \textbf{H4}. To validate \textbf{H6}, we apply the NASA-TLX~\cite{hart1988development} questionnaire before and after our experiment and correlate this with the measured engagement and language proficiency. At last, we compare the performance over time regarding pronunciation and rhythm to evaluate \textbf{H7}.

\section{Ethical and User Experience Considerations}
A multi-user, voice-based, immersive language learning application sits at the intersection of several concerns that any one of them alone would already make significant: physical embodiment, social vulnerability around speaking a language one does not yet command, biometric voice data, cloud-based AI inference, and the involvement of generative models in pedagogical content. In this section we reflect on the user experience and ethical implications of the proposed design. We begin with user experience considerations specific to immersive multi-user voice interaction, then discuss accessibility, and finally address the ethical risks and mitigations that the design must respect.

\subsection{User Experience}
Designing the user experience of an immersive multi-user VR language learning application requires attention to three properties associated with the activity that compound each other inside a head-mounted display (HMD): it is voice-centred, cognitively demanding, and physically embodied. 

\textbf{Session length} is the first design consideration. Voice-based interaction in VR produces both vocal fatigue and the well-documented physical fatigue of extended VR use \cite{ioannou2020virtual,iskander2018review}, including eye strain \cite{hirzle2022understanding}, and neck and arm strain due to sustained hand gestures \cite{pfeuffer2024design,rolff2025hands}. The optimal duration for comfortable duration with continuous VR use depends on the user's characteristics. However, studies demonstrate that benefits might decline after 45 minutes of VR duration \cite{ratan2025time}, and our design respects this convergence by structuring each quest to fit within an approximate time window \cite{hirzleUnderstandingAddressing2022,wangAssessmenteye2019,kourtesisValidationVirtual2019,hincapie-ramosConsumedendurance2014}. Prompts during the game will be placed to encourage participants to remove the headset and rest between quests, recommending a break interval between consecutive sessions rather than allowing extended use.

\textbf{Visual comfort and readability} are also extremely important, as text rendered at fixed and small sizes in VR can be unreadable for participants with mild visual impairments and uncomfortable for any participant during a long reading period \cite{gabel2023immersive,dinglerVRReading2018,buttnerinfluencetext2020}. Our design uses voice as the primary delivery channel for narrative and dialogue content, reserving on-screen text for short, high-importance elements such as spell incantations and quest objectives. Text size, contrast, and rendering distance are user-adjustable to improve readability.

The \textbf{voice dimension} introduces additional requirements specific to multi-user voice applications. Participants require detailed control over the social audio environment: the ability to adjust the volume of individual users \cite{wadley2015voice}, to mute the shared channel temporarily for individual and private practice, and to signal to other participants when they are practicing rather than speaking in-character. To mitigate issues associated with this subject, a non-verbal "practicing" gesture is visible to other players, reducing the social pressure of being overheard during difficult pronunciation attempts. Moreover, a private practice mode, available at any point, allows the participant to step out of the shared environment and rehearse spells without any other player hearing them. To avoid online harassment and toxicity, a future application will include a voice toxicity detection model \cite{nandwana2024voice}.

Considering \textbf{failure design}, it is extremely difficult to design a language-learning user experience product because speaking aloud in a language from a beginner perspective can produce high anxiety levels. Thus, a poor design will produce shame and rapid disengagement \cite{macintyre1998conceptualizing}. Our design follows previous research \cite{juul2013art,gee2005learning,foch2022game,powers2021failure} on the principle that failure should feel magical rather than punishing: a mispronounced incantation produces \textit{wild magic} -- a visual effect with a NPC reaction -- rather than a binary failure indicator. This preserves the fantasy framing while converting unsuccessful attempts into a moment of play rather than a moment of evaluation. 

\subsection{Accessibility}

Accessibility is both a design priority and a structural limitation that needs to be explicitly considered rather than minimized. Therefore, the application incorporates standard VR accessibility features, including configurable controller input mappings \cite{burdea1996multimodal}, multiple locomotion options \cite{steinicke2016being,freiwald2022continuity}, vignette comfort settings \cite{fernandes2016combating,saredakisFactorsAssociated2020}, highlighting of interactive objects, and user-adjustable text sizing. Regarding spell-casting gestures, its performance can include hand tracking or controller-based input, allowing participants to choose the modality that suits their physical comfort and the device they have available.
Notwithstanding, the voice-centric design of the spell-casting mechanic fundamentally excludes deaf, mute, and speech-impaired participants, while the visual design excludes blind and visually impaired participants. These exclusions are not consequences of careless design, as they follow the core mechanics and the choice to evaluate pronunciation and rhythm as gameplay variables. As this study is still in its preliminary form, we do not claim to have solved these accessibility issues. Yet, we understand the importance of fully considering these aspects in future work, making the limitation visible in any future deployment so that participants can make informed decisions about the system that fits their needs.

\subsection{Ethical Implications and Risks}
The ethical considerations raised by the proposed design span physical safety, psychological well-being, data protection, the integrity of AI-generated content, algorithmic bias, and broader societal impact.
\paragraph{\textbf{Physical safety:}} Immersive VR carries known risks of motion sickness, eye strain, and disorientation \cite{kennedySimulatorSickness1993,saredakisFactorsAssociated2020}. A small subset of users may experience photosensitive seizures or other neurological responses to immersive virtual environments. Participants in our study will be pre-screened for relevant contraindications during informed consent while being informed of these risks and instructed to discontinue the session immediately if they experience any adverse effects. A clearly visible exit element is available at all times.
\paragraph{\textbf{Psychological well-being:}} Language learners practicing aloud are in a state of social vulnerability \cite{horwitzForeignLanguage1986} -- specifically certain participant profiles that suffer from anxiety, com\-mu\-ni\-ca\-tion-related anxiety, or educational difficulties around language learning \cite{macintyre1998conceptualizing}. Thus, spell-casting mechanics can be presented as distressing rather than playful and pedagogical. We develop visual effects to mitigate insecurities in language learning, transforming the educational experience to be more game-like instead of competitive or punitive. 
\paragraph{\textbf{Data protection:}} Voice is classified as biometric data under the GDPR \cite{RegulationEU2016,nautschGDPRSpeech2019}, while cloud-based speech recognition and large language model services raise privacy concerns due to how voice data can carry rich incidental information beyond its intended use \cite{krogerPrivacyImplications2020}. Therefore, we do not store raw audio. Speech is transmitted to scoring services, processed, and immediately discarded; the resulting scores that are stored are linked to anonymous participant identities rather than names. Furthermore, participants are informed that voice data is processed by third-party services such as Microsoft Azure and OpenAI. Participants hold the right to withdraw and have associated local data deleted, although we are transparent that data already processed by third-party services may not be fully recoverable from these providers. 
\paragraph{\textbf{Generative AI content and hallucination:}} LLM-driven NPC dialogue introduces the risk of hallucinated cultural facts \cite{huang2025survey}, language that is grammatically correct but socially inappropriate \cite{ma2025pragmatics,solidjonov2026pragmatic}, and content that participants might absorb as authoritative or overconfident \cite{sun2025large}. We aim to mitigate this through three complementary measures: (i) cultural and linguistic content is reviewed by language experts during the co-development phase and core vocabulary and phrase is curated rather than generated; (ii) LLM dialogue is constrained through structured prompt templates and safety filters that reject inappropriate outputs; and (iii), participants are explicitly informed in the introduction that NPC responses are AI-generated and should not be treated as authoritative cultural or linguistic truth. We are aware that LLMs produce false but plausible content; thus, understanding how to mitigate these issues within a fantasy environment must be seriously considered. 
\paragraph{\textbf{Algorithmic bias in pronunciation scoring:}} Speech recognition and pronunciation assessment models are predominantly trained on prestige varieties of major languages and WEIRD (Western, Educated, Industrialized, Rich, and Democratic) \cite{sturm2015weird} populations and have been documented to score regional accents, second-language accents, and non-prototypical voices lower than prestige variants \cite{markl2022language}. This is a structural bias that affects both the validity of our pronunciation measures and, more importantly, the message it sends to learners about whose voices are heard as correct. We mitigate this by allowing participants to select among accent varieties of their target language and by framing feedback as proximity to a chosen variety rather than as binary correctness. 
\paragraph{\textbf{Accidental findings:}} The sensors used in this study can incidentally reveal information that goes beyond what we intend to measure. Sustained voice analysis during a session may surface patterns suggestive of speech, hearing, or other medical conditions \cite{krogerPrivacyImplications2020}; if eye tracking is enabled on the headset, gaze data can similarly reveal patterns associated with neurological, ophthalmological, attentional, or other conditions of which the participant may be unaware \cite{krogerWhatDoes2020,kourtesisValidationVirtual2019}. We are not in a position to diagnose, and we make no clinical claims on the basis of either modality. We mitigate the risks following three ways: (i) we collect only the data necessary to evaluate the learning outcomes specified in our research questions, not performing exploratory analyses on voice or gaze signals beyond the intended purpose; (ii) participants are informed during the consent process of the kinds of data captured by the headset and how the data is used; (iii) if patterns suggesting concern emerge during a session -- for example, sustained difficulties with articulation, irregular gaze behaviour, or signs of physical discomfort -- we follow a standard procedure of recommending the participant consult a qualified professional, without retaining or further analysing the relevant data and without communicating any specific clinical interpretation. 
\paragraph{\textbf{Societal impact:}} A capable AI-supported language learning system may, over time, contribute to the displacement of human language teachers, with both positive consequences (broader access for learners who cannot otherwise afford formal instruction) and negative consequences (loss of expertise, employment, and the cultural transmission that human teachers uniquely provide) \cite{selwynShouldrobots2019}. We acknowledge this tension with the understanding that future research needs to be addressed. We believe well-designed AI tools should complement rather than replace human teachers, while cultural and pedagogical guidance that human teachers provide is something our system explicitly relies on rather than seeking to displace.

\section{Discussion}
In this paper we have proposed a multi-user immersive VR language learning system in which players collaborate to solve fantasy role-playing fantasy quests while each learning their own chosen target languages with vocabulary, grammar, pronunciation, and rhythm embedded directly in the game mechanics (such as spell-casting). The contribution is presented as conceptual, considering the design of the platform through a human, ethical, and technological perspective.

In what follows, we reflect on the limitations and future work of the proposal. 

\paragraph{\textbf{Limitations:}} Several limitations are worth naming. The voice and vision mechanics exclude deaf, mute, blind, significantly speech-impaired participants, and vulnerable populations from the initial study; both choices are deliberate but constrain generalizability. Furthermore, the current state-of-the-art in LLMs cannot guarantee correct outputs or optimal language learning strategies. It still requires language teachers to be actively engaged in the learning process.
Another limitation is the cultural depiction inside a fantasy-themed RPG game belonging to the learned language. As each player is free to choose their language, the other players either have to learn the cultural signals depicted from the game characters for one particular player, or they have to be translated into their chosen culture. Here, more research is required to represent different cultures accurately at the same time while still being beneficial for language learning.

\paragraph{\textbf{Future Work:}}
The immediate next steps follow naturally from what this paper proposes: implementing the prototype, conducting the Phase I co-design with language experts, and running a small-scale feasibility pilot. Beyond these preliminary steps, several directions deserve dedicated research. For instance, culturally specific settings alongside the fantasy framing require further research to comprehend how cultural immersion contributes to retention and motivation (for a brief discussion see~\autoref{sec:conceptual approach}). These elements require a proper evaluation to define subsequent refinement and iterations. Beyond this, the impact of VR on language learning improvement deserves more robust and dedicated research, particularly within a gamified environment.\\

Finally, as this is a multifaceted research topic, we believe this paper has potential for laying the foundation for further research on the topic of simultaneous multi-user multi-language learning. We would like to emphasize the possibility of several studies that look into different aspects of multi-language learning in multi-user VR co-play experiences.






\section*{Disclosure of Generative AI use}
The teaser image in Figure~\ref{fig:teaser} is a conceptual rendering of our proposed VR experience and was generated using generative AI. Generative AI was also used to rephrase sentences and correct grammatical and spelling mistakes.

\begin{acks}
The authors received funding from the European University Alliance for Global Health (EUGLOH) and the German Research Foundation (DFG) EYERES (LA 952/11), DFG EXC 2176/1 - 390893254.
\end{acks}

\bibliographystyle{ACM-Reference-Format}
\bibliography{references}



\end{document}